# Implementation Of MNIST Dataset Learning Using Analog Circuit

Minjae Kim

*Abstract*—There have been many attempts to implement neural networks in the analog circuit. Most of them had a lot of input terms, and most studies implemented neural networks in the analog circuit through a circuit simulation program called Spice to avoid the need to design chips at a high cost and implement circuits directly to input them. In this study, we will implement neural networks using a capacitor and diode and use microcontrollers (Arduino Mega 2560 R3 boards) to drive real-world models and analyze the results.

## I. Introduction

DEEP learning technology is being used in various fields. It is used in fields such as line detection in autonomous vehicles. However, training a neural network requires a significant amount of power and time. Especially, In the case of Resnet, 1.7M parameters were used( He, K., et al., 2016). It is indeed true that tasks requiring such high performance are often hindered by the limitations of time and GPU capabilities on personal computers( Strubell, E., et al., 2020). The enormous cost and time required to train the neural network hinder the development of the field. Therefore, for the development of deep-learning technology, it is necessary to reduce the time and cost of training a model. In this paper, it was determined that analog computing could be the answer. There were many researchers who thought of this, and in fact, efforts were made to develop deep-learning models using analog circuits, but they tended to be vulnerable to Noise because they used analog formats rather than discrete digital formats. Since the resistance value was used as the weight of the matrix, the results were different due to the error rate that the resistance value inevitably had (Mead, C. and M. Ismail, 1989). Also excessive power consumption, and large area occupied by discrete resistors and operational amplifiers. Resistance of discrete fixed resistors that cannot be easily adjusted or easily controlled. As a result, there was a problem that could not be used for learning and could not be reused for reclamation in the solution of the new task (Zurada, J. M., 1992). In this study, we will use a new structure instead of the existing Matrix-based deep-learning structure to eliminate instability from devices in analog neural network circuits and build circuits that are superior in speed, power consumption, and price to deep-learning model training courses using GPU or TPU. We will also use this structure to learn MNIST data and analyze performance.

### A. Analysis of existing research.

In the case of existing studies, a neural network structure was implemented as an electronic circuit and results were obtained using the value of the current flowing through the voltage. For example, the neural network structure was implemented and used as an electronic circuit as shown in Figure 2. On the right side of the figure, it consists of an inhibitory interconnection between output neurons. These inhibitory neurons suppress each other so that only neurons with strong output survive (Graf, H. P. and L. D. Jackel., 1989). The limitation of these circuits is that the resistance value is encapsulated. Resistance values cannot be easily adjusted or controlled, resulting in problems that cannot be used for learning (Zurada, J. M., 1992).

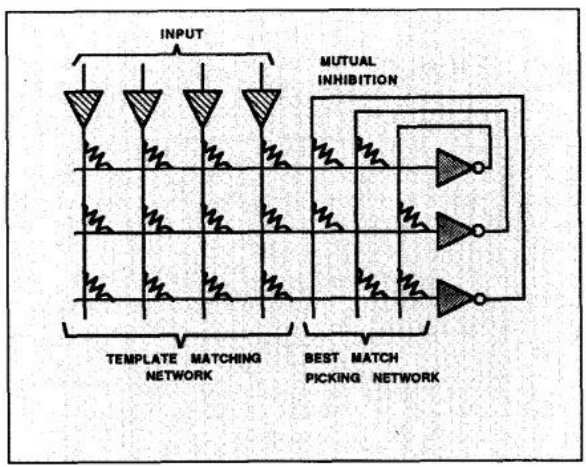

Fig. 1.[6] This figure shows the diagram of the previously used analog neural network.

*B. Existing Research 1. Analog Neural Network*

Analog neural networks receive voltage, and current, as continuous positive input. Unlike most deep-learning models that we use in computers, the brain, especially neuron cells, works analogically. Many artificial neural network LSIs operate in an analog way, including various operators such as addition, multiplication, and nonlinear operators within a single chip, and have speed advantages over digital methods. In pure analog circuits, the main problem is the achievement of analog memory (Chong, C. P., et al., 1992).Especially, saving analog value is a problem that has not yet been solved. The temporary method is to use a capacitor as memory. The method is remembered as temporary memory in the capacitor because it can be achieved in general purpose CMOS Process. However, terminology and digital memory, will also be required if data values persist for a long time. In this case, the D/A and A/D conversions cause overhead problems. (Kawaguchi, M., et al. ,2015).

*C. Pulsed Neural Network*

The pulsed neural network is another type of analog neural network. When processing time series data, pulsed neural network models have good advantages. Moreover, this network can maintain connection weights after the learning process. In addition, the reason for using capacitors in the learning circuit is that the circuit takes a long time to operate. The pulse neuron model represents the output value by the firing probability of the neuron. This characteristic is the cause of the slow response speed of the pulsed neural network.( Kawaguchi, M., et al. , 2015).

## II. EXPERIMENTAL SETUP

In this paper, we will construct a circuit that solves the noise problem of analog deep-learning circuits and analyze it in terms of accuracy, operating time, and power consumption using the MNIST dataset. The study consists of the following four processes. 1. Classification task for the MNIST dataset 2. Circuit assembly using modules proposed earlier 3. Use the completed circuit and MCU (Microcontroller {Arduino Mega 2560 Board}) to proceed with the learning. 4. Comparison of MNIST dataset's learning results and performance in Python using Numpy (accuracy, operating speed, power consumption)

Through the above process, we confirm whether the circuit we designed works well as a neural network, and analyze the accuracy, operating time, and power consumption of the circuit.

*A. Preprocessing For Dataset*

1. The MNIST dataset to be used as input data[9] consists of 60000X785. The first column represents the number corresponding to the figure. The remaining 784 columns represent the input image. Convert the 1X784 matrix to 28X28 Matrix, change it from 28X28 to 30X30 via zero padding, and resize the matrix to 5X5 Size.

2. Change the form of the matrix (5X5 matrix to 1X25 matrix). These processes are implemented throughout the MNIST dataset. After the process, 60000X25 arrays are formed.

3. Send data values in an excel file into a text file and remove any gaps in the text file.

4. Paste the MNIST dataset's Result values (meaning the values of the first columns of CSV in step 1) into a text file and save them.

5. Copy the text file generated in number 3 and the text file generated in number 4 and save it to the Transcend 2GB micro SD card.[10]

*B. Circuit generation*

It creates a circuit to proceed with the deep-learning process. The circuit consists largely of four types of submodules.

*1) Module 1*

Fig. 2 shows a schematic diagram of Module 1. $R_1$ and $R_3$ are 330 ohms and $R_2$ are 10 ohms. $C_1$ is 10uF. For diodes, 1N4148 was used. In the case of this circuit, the value of the resistance applied to the cathode varies depending on the time the voltage is applied. The varying resistance value may vary the amount of current outputted with respect to the input voltage. In addition, the resistance value changes only when the voltage of the input terminal is higher than the output terminal, and if not, the resistance value is stored. Circuit design in this way is not a reckless challenge. As shown in Fig. 3, attempts have been made in the past to adjust and store the resistance value through the Capacitor.

*2) Module 2*

The overall structure is the same as Module 1, but the value of $C_1$ has been changed to 1uF. Module 1 was saturated first by placing a difference in the value of $C_1$.

*3) Module 3 And Input Process*

It serves to collect the input values of Module 1.
By combining the completed modules, we created a processor that performs deep-learning operations. The diagram of the processor is shown in Fig. 5.

*4) Proof of non-linearity of Module 1 and Module 3*

The multilayer analog circuit configured as shown in Fig. 6 must be nonlinear to be meaningful as a multilayer analog circuit. If the modules (module 1, module 2, and module 3) constituting each layer are linear, an analog circuit composed of multiple layers can be equivalently converted into an analog circuit composed of a single layer. Therefore, we want to prove that the circuit we construct is nonlinear. Since module 3 is a layer composed of diodes, there is a nonlinearity between the negative and positive voltage ranges. However, there may be cases where the circuit we build operates only in a positive voltage range. It is intended to show that nonlinearity is established even when operating in a positive voltage range. To do so, we will first find the circuit equation of module 1. Since the equation is developed on the premise of a positive voltage range, the diode is considered an ideal diode? The equation will be developed in the replacement of the diode with a wire.

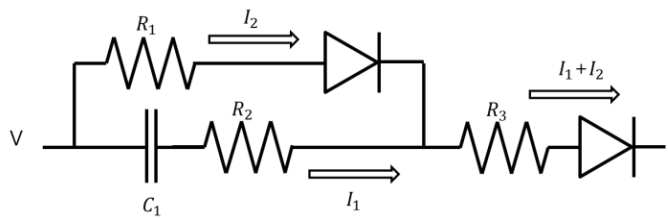

Fig. 2. This figure shows the schematic diagram of module1. The resistance value varies depending on the size and time of the input voltage.

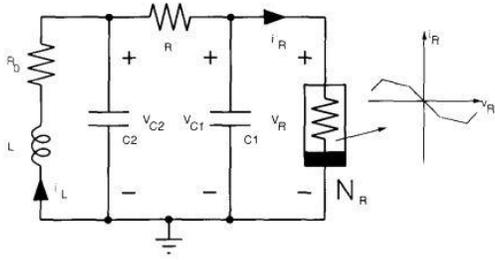

Fig. 3 [10]. This figure shows that there have been previous attempts to change and store the value of weight using a capacitor.

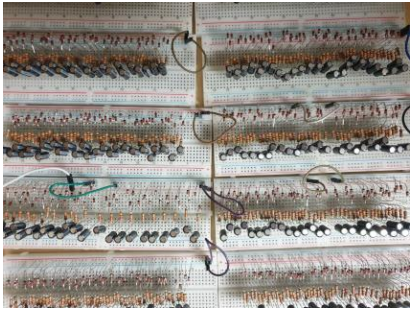

Fig. 4. This figure shows Module 1 as it really is.

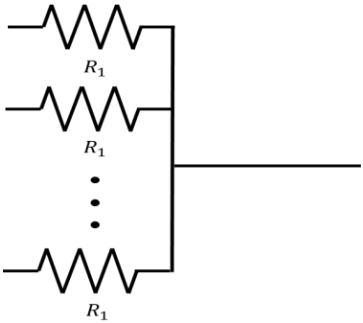

Fig. 5. This figure shows the schematic diagram of module3.

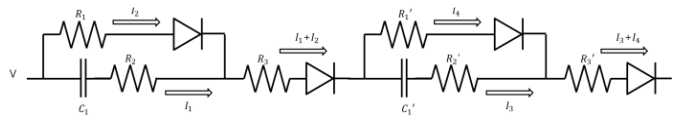

Fig. 6. This figure is a scheme diagram to show that module1 and module3 have nonlinearity.

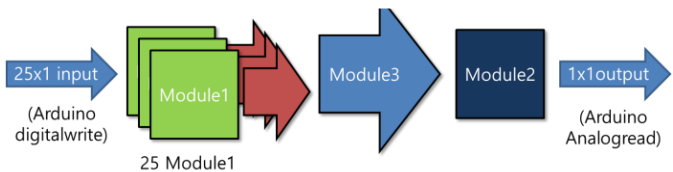

Fig. 7. This figure shows how the proposed circuit works.

The following formula is established according to Kirchhoff's law.

**Equation 1,2**

$$V - R_3(I_1 + I_2) - R_1 I_2 = 0$$
$$R_1 I_2 = R_2 I_1 + \frac{\int I_1 dt}{C_1}$$

When $I_2$ in Equation 2 is substituted for Equation 1, the following equation is obtained.

**Equation 3**

$$V = \left(\frac{R_1 R_3 + R_1 R_2 + R_2 R_3}{R_1}\right) I_1 + \left(\frac{R_3 + R_1}{R_1 C_1}\right) \int I_1 dt$$

What is important in this equation is that the circuit equation of Module 1 consists of the integral term of $I_1$ and the constant term multiplied by $I_1$. Now, it is time to look at whether the circuit combined with modules 1,2, and 3 can be represented by the equivalent conversion of Module 1. As before, we will calculate the circuit equation assuming that it is an ideal diode and a forward voltage is applied. As before, the following circuit equations are established according to Kirchhoff's law.

**Equation 4**

$$V - R_3'(I_3 + I_4) - R_1' I_4 = 0$$

**Equation 5**

$$R_1' I_4 = R_2' I_3 + \frac{\int I_3 dt}{C_1'}$$

**Equation 6**

$$I_1 + I_2 = I_3 + I_4$$

By combining Equations 1, 2, 4, 5, and 6, the following equations can be obtained.

**Equation 7**

$$V - (R_3 + R_3')\left(I_1 + \frac{R_2}{R_1} I_1 + \frac{\int I_1 dt}{R_1 C_1}\right) - R_2 I_1 - \frac{\int I_1 dt}{C_1} - R_2' I_3 - \frac{\int I_3 dt}{C_1} = 0$$

**Equation 8**

$$I_1 + \frac{R_2}{R_1} I_1 + \frac{\int I_1 dt}{R_1 C_1} = I_3 + \frac{R_2'}{R_1'} I_3 + \frac{\int I_3 dt}{R_1 C_1'}$$

If $I_3$ is completely replaced in Equation 7 through Equation 8, the corresponding circuit with a two-layer structure can be expressed using Module 1. However, $I_3 + \frac{R_2'}{R_1'} I_3 + \frac{\int I_3 dt}{R_1 C_1'}$ of Equation 8 and $R_2' I_3 + \frac{\int I_3 dt}{C_1}$ of Equation 7 are not professional, so $I_3$ or integral terms of Equation7 inevitably remain.

Therefore, it may be seen that the multi-layered circuit cannot be expressed as a single-layered equivalent circuit. The circuit design simulation results are as follows. The simulation used falstad.com .

The yellow graph represents the first circuit in Figure, and the red graph represents the second circuit on Figure.

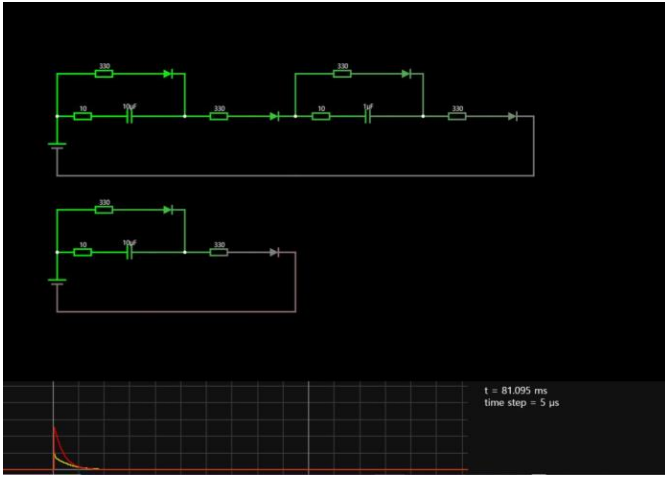

Fig. 8. This figure shows the result of simulating the circuit of Figure 6.

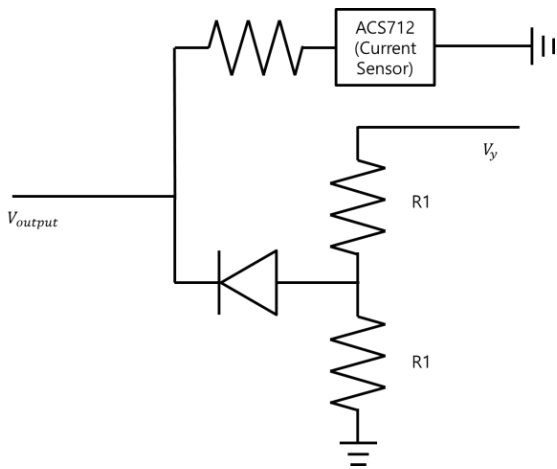

Fig. 9. This figure shows the schematic diagram of module 1. Learning is conducted through voltage and vy values output through module 1, module 2, and module 3, and the values output through module 1, module 2, and module 3 are read through the ACS712 module.

*5) Module 4*

The value output through module 1-module 3-module 2 and the value output from module 4 meet with a diode interposed there between. If the potential of point A is higher than the potential of point B (y(target)=0, y(real)=1), the reverse electromotive force is applied to module 1-module 3-module 2 at point B, reducing the speed at which the resistance increases. The weight (resistance) values are adjusted in this way.

*C. Performance Comparison*

7. The SD card, which stores three times of data, reads data from Arduino through the Arduino SD card reader module. The read data is input to the processor through the Arduino Digitalwrite function as shown in the figure. The results processed through the Processor were input to the computer through the Arduino Current Sensor Module (ACS712) Arduino Analogread function. At this time, Arduino 2560 board was used. In addition, the DigitalWriteFast package was used to speed up the execution of digitalwrite functions. With the DigitalWriteFast package, the output speed changes from 5.8us to 128ns.[11]
8. Verify that the entire process works well as a deep-learning module.
9. The process of 1-8 is repeated to obtain the validity of the results. At this time, the results are analyzed through the confusion matrix.

### III. RESULTS

There were some errors in the accuracy. The reason was that it took about 4 minutes to obtain one result, and the number of awards in Module 1 and Module 2 was very small compared to the runtime of 10 minutes, so it is analyzed that the Capacitors in Module 1 and Module 2 were saturated before the learning was completed. In terms of Operation Time, it was also somewhat lacking: the code operated in Colab was completed in 14 seconds, while our model using Arduino and self-made circuits took approximately 40 minutes. However, this is due to clock problems and memory problems with Arduino processors. For example, the clock of the Intel i7-1360 used in computers is 5GHZ, while the clock of the Arduino 2560 R3-compatible module is only 16MHZ. It is not possible to compare exactly, but there is a speed difference of about 300 times. Our model consists of 10 iterations of 1X1 OutPut ,unlike the conventional one with an output of 10X1. With more materials, we could have built a 10X1 Output model, which leads to a 10x speed increase. In addition, Python used to extract and use all datasets in memory, but in the case of the model we designed, it was slower than the method using memory because the txt file was read from the SD card at any time through SPI communication.
The graph for the model operating time ratio is shown in figure 11.

![Arduino code screenshot]

Fig. 10. This figure shows the Arduino code for running this model.

|   | 0 | 1 | 2 | 3 | 4 | 5 | 6 | 7 | 8 | 9 |
|---|---|---|---|---|---|---|---|---|---|---|
| 0 | -1.1 | -0.5 | -0.3 | 0.6 | 6E-14 | 0.5 | -0.2 | 0.2 | 1.3 | 1.2 |
| 1 | -1.5 | -0.9 | -0.8 | 0.8 | 1 | -1.6 | 1 | -0.6 | 0.2 | 2.2 |
| 2 | -0.2 | -1.6 | -0.3 | -0.2 | 0.6 | 0 | 1.2 | 0.6 | -0.3 | 0.9 |
| 3 | -0.7 | -0.1 | 0.7 | 1.3 | 1.5 | 2.3 | 0.5 | 0.3 | 0.6 | -1.6 |
| 4 | -0.6 | 0.1 | -1.9 | 0.3 | 0 | -0.9 | -0.2 | 6E-14 | 0.9 | 0.3 |
| 5 | 2.7 | 0.1 | -1.9 | -1.7 | -3.1 | -2.1 | -0.7 | -0.7 | 2.3 | -1 |
| 6 | -1.8 | 1.7 | -1.1 | -1.5 | -0.3 | -0.6 | 6E-14 | 0.8 | 1.5 | -1.3 |
| 7 | -3.2 | -0.4 | 2.7 | -1.1 | -1.6 | 0.6 | 2.4 | 0.5 | -0.2 | 0.2 |
| 8 | -0.8 | -0.2 | 1.4 | -1.4 | 1 | -1.4 | 0.5 | 0.3 | 3.2 | 2 |
| 9 | -0.7 | 0.7 | 0.3 | -1 | -0.5 | 0.2 | 1.3 | -0.2 | -0.3 | -1.9 |

Fig. 11.
This figure displays the output results for the test dataset of a model trained on the MNIST dataset. Green cells indicate matching expected and output values, apricot cells represent cases where the top 3 expected values match the output value, blue cells signify discrepancies between expected and output values, and red cells highlight the lowest value when the expected and output values do not match.

TABLE I
RESULT COMPARISON

|  | Numpy Based Colab[13] | Ours |
|---|---|---|
| Operation Time | 14s | 37 min 39 s |
| Accuracy | 0.9 | 0.3(0.7) |

Comparison of numpy-based deep learning results with our model in pc. The results in parentheses reflect the Top 3 score. Accuracy was measured as (the number of expected and actual matches/the total number).

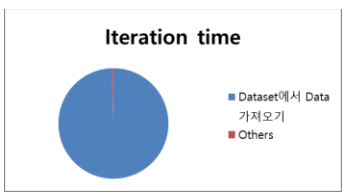

Fig. 12. Iteration Time Comparison

## IV. DISCUSSION

Compared to previous studies, it has an advantage in that it does not use fixed resistance as a weight, but optimizes the weight on its own. In the Analog Neural Network, the goal was to reduce the error produced by the error in the magnitude of the resistor, but it was not successful in reducing the error. The reason is that the current at the OUTPUT end is strongly affected by the number of voltages at the INPUT end. To solve this problem, we tried to increase accuracy by inputting the circuit before learning and setting the output to Baseline and MNIST-Classification through the difference, but there was an error. The second cause is the measurement uncertainty of the ACS712 model (current sensor). The following document identifies measurement performance issues with the ACS712. The last problem is the storage and output structure of the dataset. According to the draft, we were able to store dataset in Memory and load it at a high speed, which means that the Capacitor in module 1 was not fully charged until the learning was completed. However, using SPI communication using SD cards took 0.4 seconds to load dataset's 1row. This model has a saturation problem in which the Capacitor (memory) is fully charged in one termination. To solve this problem, we had to change the structure of the model from a model that could be charged and discharged to a model that could only be charged. However, the model is still competitive enough in terms of speed, price, and energy consumption. If the output structure of the model is changed from 1X1 to 10X1, the output time is 1/10, and if the data set is accessed through memory rather than through SPI communication using another microcontroller or mini PC (Raspberry Pi), the execution speed can be optimized up to 2GHZ. The advantage of this model is that the time according to the input size of the model follows the n-time complexity. Therefore, there is clearly an advantage in terms of learning time for deep learning for large-sized inputs.
In the future, we will measure Baseline accurately with more accurate current sensors, use a memory approach, and improve our model in a good way.

## V. CONCLUSION

In this paper, we conducted the process of learning MNIST data through self-produced circuits.The learning was conducted in a different way from the analog circuits for deep-learning, which were immersed in the multiplication operation of the matrix, and showed some freedom in the noise of the elements forming the circuit. In addition, it showed less power consumption than analog deep-learning circuits using conventional passive devices. The operating time may seem a bit slow, but this is due to the fact that the Arduino 2560 R3 board, a microcontroller used, provides a clock of 16MHZ, which is not as good as the computer CPU, and that board does not have enough memory to quickly output Dataset output is impossible memory. With a controller that provides enough memory and a clock, it's going to be even better.

REFERENCES

[1] He, K., et al. (2016). Deep residual learning for image recognition. Proceedings of the IEEE conference on computer vision and pattern recognition. Letter Symbols for Quantities, ANSI Standard Y10.5-1968.

[2] Strubell, E., et al. (2020). Energy and policy considerations for modern deep-learning research. Proceedings of the AAAI conference on artificial intelligence.

[3] Mead, C. and M. Ismail (1989). Analog VLSI implementation of neural systems, Springer Science & Business Media.

[4] Zurada, J. M. (1992). "Analog implementation of neural networks." IEEE Circuits and Devices Magazine 8(5): 36-41.

[5] Graf, H. P. and L. D. Jackel (1989). "Analog electronic neural network circuits." IEEE Circuits and Devices Magazine 5(4): 44-49.

[6] Graf, H. P. and L. D. Jackel (1989). "Analog electronic neural network circuits." IEEE Circuits and Devices Magazine 5(4): 44-49.

[7] Zurada, J. M. (1992). "Analog implementation of neural networks." IEEE Circuits and Devices Magazine 8(5): 36-41.

[8] Chong, C. P., et al. (1992). "Image-motion detection using analog VLSI." IEEE Journal of Solid-State Circuits 27(1): 93-96.

[9] MNIST in CSV. (2018, May 19). Kaggle. https://www.kaggle.com/datasets/oddrationale/mnist-in-csv

[10] Roska, T. and L. O. Chua (1993). "The CNN universal machine: an analogic array computer." IEEE Transactions on Circuits and Systems II: Analog and Digital Signal Processing 40(3): 163-173.

[11] GitHub - ArminJo/digitalWriteFast: Arduino library for faster and smaller digitalWrite(), digitalRead() and pinMode() functions using direct port manipulation for constant pin numbers.

[12] Kawaguchi, M., et al. (2015). Analog neural circuit with switched capacitor and design of deep learning model. 2015 3rd International Conference on Applied Computing and Information Technology/2nd International Conference on Computational Science and Intelligence, IEEE.

[13] https://colab.research.google.com/drive/1Grr0hjRPfGPk7Ti6Dy8L1tU7xtm83eQa?usp=sharing (reformed from https://github.com/makeyourownneuralnetwork/makeyourownneuralnetwork.git)